\documentclass[aps, reprint, nofootinbib]{revtex4-2}

\usepackage{amsfonts}
\usepackage{amsmath}
\usepackage{graphicx}
\usepackage{algorithm}
\usepackage{algpseudocode}
\usepackage{enumitem}
\usepackage{listings}
\usepackage{upquote}
\usepackage{float}

\newcommand{\rnorm}{\hat{\rho}}
 
\newcommand{\phibdot}{\dot{\Phi}_b}
\newcommand{\phib}{\Phi_b}

\begin{document}

\title{TORAX: A Fast and Differentiable Tokamak Transport Simulator in JAX}

\author{Jonathan Citrin}
\email{citrin@google.com}
\affiliation{Google DeepMind, London, UK}
\author{Ian Goodfellow}
\affiliation{Google DeepMind, London, UK}
\author{Akhil Raju}
\affiliation{Google DeepMind, London, UK}
\author{Jeremy Chen}
\affiliation{Google DeepMind, London, UK}
\author{Jonas Degrave}
\affiliation{Google DeepMind, London, UK}
\author{Craig Donner}
\affiliation{Google DeepMind, London, UK}
\author{Federico Felici}
\affiliation{Google DeepMind, London, UK}
\author{Philippe Hamel}
\affiliation{Google DeepMind, London, UK}
\author{Andrea Huber}
\affiliation{Google DeepMind, London, UK}
\author{Dmitry Nikulin}
\affiliation{Google DeepMind, London, UK}
\author{David Pfau}
\affiliation{Google DeepMind, London, UK}
\author{Brendan Tracey}
\affiliation{Google DeepMind, London, UK}
\author{Martin Riedmiller}
\affiliation{Google DeepMind, London, UK}
\author{Pushmeet Kohli}
\affiliation{Google DeepMind, London, UK}


\begin{abstract}
We present TORAX\footnote{https://github.com/google-deepmind/torax{\newline}https:/torax.readthedocs.io}, a new, open-source, differentiable tokamak core transport simulator implemented in Python using the JAX framework. TORAX solves the coupled equations for ion heat transport, electron heat transport, particle transport, and current diffusion, incorporating modular physics-based and ML models. JAX's just-in-time compilation ensures fast runtimes, while its automatic differentiation capability enables gradient-based optimization workflows and simplifies the use of Jacobian-based PDE solvers. Coupling to ML-surrogates of physics models is greatly facilitated by JAX's intrinsic support for neural network development and inference. TORAX is verified against the established RAPTOR code, demonstrating agreement in simulated plasma profiles. TORAX provides a powerful and versatile tool for accelerating research in tokamak scenario modeling, pulse design, and control.
\end{abstract}


\maketitle

\section{Introduction}

Toroidal magnetic confinement devices such as tokamaks and stellarators are currently the leading candidates for fusion energy reactors~\cite{ongena:2016}. Accurate integrated plasma simulation is crucial for a wide range of applications, including interpretation of experiments, physics validation and understanding, extrapolation to future performance, scenario discovery and optimization, and controller design~\cite{poli:2018}. Integrated tokamak modelling is inherently multi-scale and multi-physics, with multiple orders of magnitude in spatiotemporal scales between relevant physics processes~\cite{fasoli:2016}. Therefore domain decomposition and physics reduction is key. 

Core transport simulation focuses on the plasma domain with closed nested magnetic flux surfaces, where due to fast transport along field lines, the transport problem can be reduced to a set of coupled 1D transport PDEs. The bulk of fusion power and plasma current is concentrated in the plasma core. Core simulation is thus a central component of  prediction of the (fusion) performance of a plasma, combining plasma equilibrium, transport, fuelling, and heating. 

Multiple core transport codes have been developed, including JETTO~\cite{cenacchi:1988}, ASTRA~\cite{pereverzev:1991}, PTRANSP~\cite{transp:2018}, and ETS~\cite{kalupin:2013}. These codes offer a range of physics fidelity in their constituent physics models, with higher fidelity models often coming at the expense of simulation speed. Simulations can take hours to days on a standard compute node for simulating several energy confinement times when using state-of-the-art reduced turbulence, neoclassical, and heating modules. 

However, a range of use-cases demands both fast and accurate simulations.  Crucially, differentiability is essential for gradient-based optimization and sensitivity analysis, which is typically not available in traditional transport codes.  Speed is also vital for many-query problems exploring a wide range of parameter spaces, for example for uncertainty quantification (UQ), scenario optimization, inter-shot experimental preparation, model-based controller design, and reactor design.

For optimization and control-oriented applications, dedicated codes have been developed, such as RAPTOR~\cite{felici:2012, felici:2018}, METIS~\cite{artaud:2018}, FENIX~\cite{janky:2021}, and COTSIM~\cite{morosohk:2021}. To circumvent the conflicting constraints of simulation speed and accuracy, machine learned (ML) surrogates of complex physics models has emerged as a promising technique~\cite{meneghini:2017,vandeplassche:2020,ho:2023,boyer:2019,citrin:2023genenn,joung:2023,rodriguez-fernandez:2022c,morosohk:2024}. However, all the codes listed thus far are developed in Fortran or MATLAB, which can pose challenges for (auto)-differentiation and seamless coupling to ML-surrogate models. RAPTOR stands out as a differentiable transport code, but it is not auto-differentiable, and thus relies on manually coded Jacobian calculations, which is a hindrance for extensibility, especially to ML-surrogates which are complex nested nonlinear functions.

Therefore, existing tokamak transport codes face limitations in terms of a subset of computational speed,  auto-differentiability, and integration with emerging machine learning (ML) techniques. 

Addressing these limitations, we present TORAX, a new, open-source, tokamak core transport simulator implemented in Python using the JAX framework~\cite{jax:2018github}. JAX provides automatic differentiation and just-in-time (JIT) compilation, enabling both fast runtimes and gradient-based algorithms, including optimization and nonlinear PDE solvers. Coupling to ML-surrogates of physics models is greatly facilitated by JAX's intrinsic support for neural network development and inference, allowing such surrogates to be developed in the native language of the simulator. Python's ease of use and wide adoption facilitates code extensibility, maintainability, and coupling within various workflows and to additional physics models. Finally, JAX can seamlessly execute on multiple backends including CPU and GPU, with support for vectorization and parallelization.

In this paper, we outline the TORAX model and methodology, highlighting its key initial features and capabilities. We benchmark TORAX against the established RAPTOR code, demonstrating accurate and verified performance. Following this initial release, further TORAX extensions and generalizations are planned. We conclude by outlining the development roadmap.

\section{TORAX model description}
TORAX simulates the time evolution of core plasma profiles using a coupled set of 1D transport PDEs, discretized in space and time, and solved numerically. The PDEs arise from flux-surfaced averaged moments of the underlying kinetic equations, and from Faraday's law~\cite{hinton:1976}. Their validity relies on time-scale separation of transport and turbulent dynamics, and on kinetic effects (e.g. turbulent and neoclassical dynamics) being a small perturbation to the background equilibrium distribution functions, conditions generally satisfied in the tokamak or stellarator core.

\subsection{Governing equations}
\label{sec:equations}
\begin{figure}[hbt]
    \includegraphics[width=1.0\linewidth]{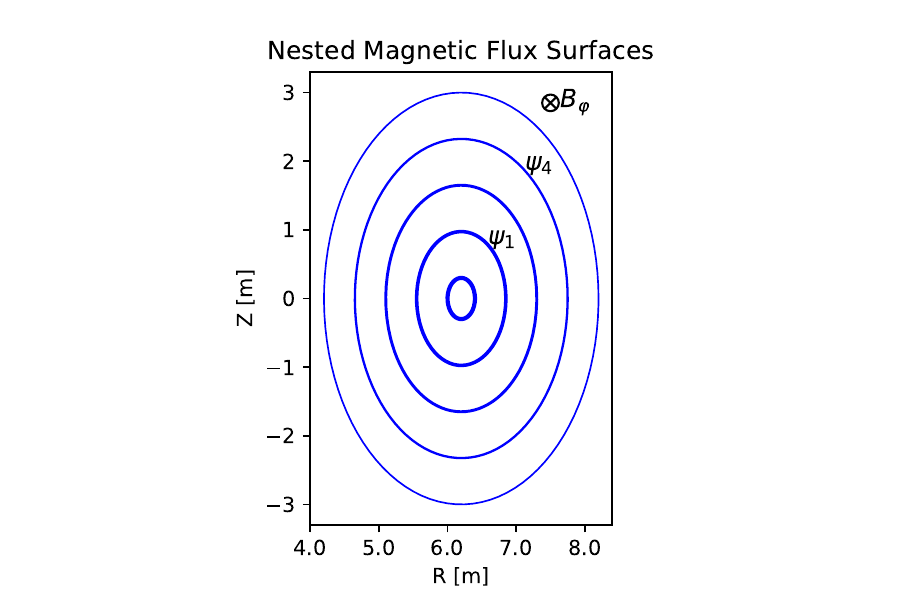}
    \caption{\footnotesize Poloidal cross section of the core region of a toroidal tokamak plasma, with closed nested magnetic flux surfaces. R is the major radius from the center of the torus, Z is the vertical direction. ITER-like magnitudes are shown. The flux surfaces are elliptical in the poloidal plane for illustrative purposes, and do not correspond to a true equilibrium calculation. The toroidal magnetic field $B_{\varphi}$ is in the direction into the page. $\psi_1$ and $\psi_4$ label two surfaces of constant poloidal flux at arbitrary locations. A discrete set of toroidal flux coordinates $\rho$ is defined through integrating the toroidal magnetic field within a discrete set of poloidal flux surfaces.}
    \label{fig:flux_surfaces}
\end{figure}

TORAX solves coupled 1D PDEs in normalized toroidal flux coordinates,  $\rnorm$, with $0 \leq \rnorm \leq 1$. $\rnorm$, is a flux surface label, being constant on a closed surface of poloidal magnetic flux $\psi$. It is defined as as $\rnorm=\sqrt{\frac{\Phi(\psi)}{\phib}}$, where $\Phi(\psi)$ is the
toroidal magnetic flux enclosed by the magnetic poloidal flux surface being labelled (see figure~\ref{fig:flux_surfaces}), and $\phib$ is the toroidal flux enclosed by the plasma core boundary, i.e. the last-closed-flux-surface.

The set of 1D PDEs being solved are the following:

\begin{itemize}
    
    \item Ion heat transport, governing the evolution of the ion temperature $T_i$.
    \begin{multline}
    \label{eq:Ti}
    \frac{3}{2} V'^{-5/3} \left(\frac{\partial }{\partial t}-\frac{\phibdot}{2\phib}\frac{\partial}{\partial\rnorm}\rnorm\right)\left[V'^{5/3} n_i T_i\right] = \\ \frac{1}{V'} \frac{\partial}{\partial \rnorm} \left[\chi_i n_i \frac{g_1}{V'} \frac{\partial T_i}{\partial \rnorm} - g_0q_i^{\mathrm{conv}}T_i\right] + Q_i
    \end{multline}

    \item Electron heat transport, governing the evolution of the electron temperature $T_e$.
    \begin{multline}
    \label{eq:Te}
    \frac{3}{2} V'^{-5/3} \left(\frac{\partial }{\partial t}-\frac{\phibdot}{2\phib}\frac{\partial}{\partial\rnorm}\rnorm\right)\left[V'^{5/3} n_e T_e\right]  = \\ \frac{1}{V'} \frac{\partial}{\partial \rnorm} \left[ \chi_e n_e \frac{g_1}{V'} \frac{\partial T_e}{\partial \rnorm} - g_0q_e^{\mathrm{conv}}T_e \right] + Q_e
    \end{multline}

    \item Electron particle transport, governing the evolution of the electron density $n_e$.
    \begin{multline}
    \label{eq:ne}
    \left(\frac{\partial}{\partial t}-\frac{\phibdot}{2\phib}\frac{\partial}{\partial\rnorm}\rnorm\right) \left[ n_e V' \right] = \\ \frac{\partial}{\partial \rnorm} \left[D_e\frac{g_1}{V'} \frac{\partial n_e}{\partial \rnorm} - g_0V_e n_e\right] + V'S_n
    \end{multline}    

    \item Current diffusion, governing the evolution of the poloidal flux $\psi$.
\begin{multline}
    \label{eq:psi}
 \frac{16 \pi^2 \sigma_{||}\mu_0 \rnorm \phib^2}{F^2}\left(\frac{\partial \psi}{\partial t}-\frac{\rnorm\phibdot}{2\phib}\frac{\partial \psi}{\partial \rnorm}\right)  = \\ \frac{\partial}{\partial \rnorm} \left( \frac{g_2 g_3}{\rnorm} \frac{\partial \psi}{\partial \rnorm} \right) - \frac{8\pi^2 V' \mu_0 \phib}{F^2} \langle \mathbf{B} \cdot \mathbf{j}_{ni} \rangle 
\end{multline}    
\end{itemize}

where $T_{i,e}$ are ion and electron temperatures, $n_{i,e}$ are ion and electron densities, and $\psi$ is poloidal flux. $\chi_{i,e}$ are ion and electron heat conductivities, $q_{i,e}^{\mathrm{conv}}$ are ion and electron heat convections, $D_e$ is electron particle diffusivity, and $V_e$ is electron particle convection. $Q_{i,e}$ are the total ion and electron heat sources, and $S_n$ is the total electron particle source. $V' \equiv dV/d\rnorm$, i.e. the flux surface volume derivative. $\sigma_{||}$ is the plasma neoclassical conductivity, and $\langle \mathbf{B} \cdot \mathbf{j}_{ni} \rangle$ is the flux-surface-averaged non-inductive current (comprised of the bootstrap current and external current drive) projected onto and multiplied by the magnetic field. $F \equiv RB_\varphi$ is the toroidal field flux function, where $R$ is major radius and $B_\varphi$ the toroidal magnetic field. $\phib$ is the toroidal flux enclosed by the last closed flux surface, and $\phibdot$ is its time derivative, non-zero with time-varying toroidal magnetic field and/or last closed flux surface shape. $\mu_0$ is the permeability of free space. The geometric quantities $g_0$, $g_1$, $g_2$ and $g_3$ are defined as follows. 

\begin{equation}
\label{eq:g0}
g_0 = \left< \left( \nabla V \right) \right> 
\end{equation}

\begin{equation}
\label{eq:g1}
g_1 = \left< \left( \nabla V \right)^2 \right> 
\end{equation}
where $\nabla V$ is the radial derivative of the plasma volume, and $\langle \rangle$ denotes flux surface averaging,
\begin{equation}
\label{eq:g2}
g_2 = \left< \frac{\left( \nabla V \right)^2}{R^2}\right> 
\end{equation}

\begin{equation}
\label{eq:g3}
g_3 = \left< \frac{1}{R^2}\right>
\end{equation}
where $R$ is the major radius along the flux surface being averaged.

The geometry terms $V'$, $g_0$, $g_1$, $g_2$ and $g_3$ are calculated from flux-surface-averaged outputs of a Grad-Shafranov equilibrium code (see section~\ref{sec:models}), either pre-calculated or coupled to TORAX within a larger workflow.

The boundary conditions are as follows. All equations have a zero-derivative boundary condition at $\rnorm=0$. The $T_i$, $T_e$, $n_e$ equations have fixed boundary conditions at $\rnorm=1$, which are user-defined. The $\psi$ equation has a Neumann (derivative) boundary condition at $\rnorm=1$, which sets the total plasma current through the relation:
\begin{equation}
    I_p = \left[\frac{\partial \psi}{\partial \rho} \frac{g_2 g_3}{\rho}\frac{R_0 J}{16\pi^4\mu_0}\right]_{LCFS}
\end{equation}

See Ref.~\cite{felici:2011} for a more comprehensive summary and derivation of the governing equations and associated geometric quantities. See Appendix~\ref{appendix:phibdot} for a breakdown in how the $\phibdot$ term leads to additional convective and source terms. Future work can extend the governing equations to include momentum transport, and multiple ion species including impurities. Details of the physics models underlying the PDE coefficients is provided in section~\ref{sec:models}.

\subsection{Spatial discretization scheme}
\label{sec:fvm}
\begin{figure}[hbt]
    \includegraphics[width=1.0\linewidth]{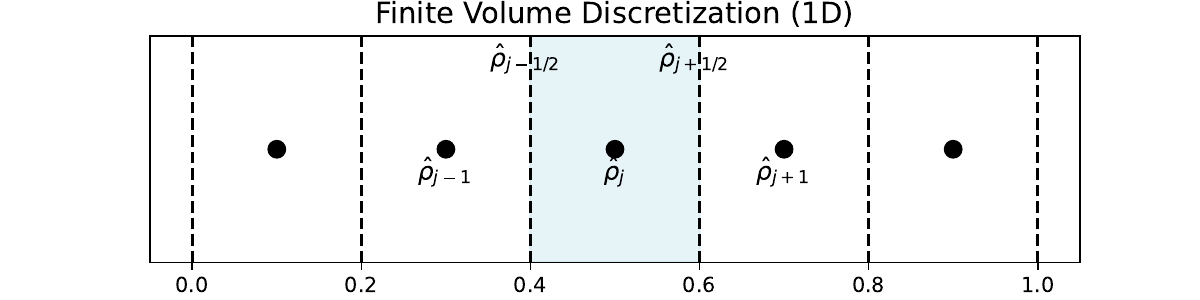}
    \caption{\footnotesize 1D spatial discretization with the finite volume method, with $N=5$ and $\rnorm=0.2$. The cell volume centered at $\rnorm=0.5$ is highlighted. Keeping with 3D FVM formalism, a cell volume is defined with $V=d\hat{\rho}A$, with arbitrary face area $A$. The corresponding cell faces are located at $\rnorm=0.4$ and $\rnorm=0.6$. In the 1D case spatial inhomogeneity is only along the $\rnorm$ direction, and the final equations are normalized by $V$, cancelling out $A$.}
    \label{figure:fvm}
\end{figure}

TORAX employs a finite volume method (FVM) to discretize the governing PDEs in space~\cite{eymard:2000}. The TORAX JAX 1D FVM package is significantly influenced by FiPy~\cite{FiPy:2009}. The 1D spatial domain, $0 \leq \rnorm \leq 1$, is divided into a uniform grid of $N$ cells, each with a width of $d \rnorm = 1/N$.  The cell centers are denoted by  $\rnorm_i$,  where  $0 = 1, 2,..., N-1$, and the $N+1$ cell faces are located at $\rnorm_{i\pm1/2}$.  Both $\rnorm=0$ and $\rnorm=1$ are on the face grid. TORAX is currently restricted to a uniformly spaced grid, with generalization to a non-uniform grid left for future development. See figure~\ref{figure:fvm} for an illustration of the spatial grid.

The FVM approach involves integrating each PDE over a control volume (in this case, a cell) and applying the divergence theorem to convert volume integrals to surface integrals. This leads to a system of algebraic equations for the cell-averaged values of the plasma profiles. 

For a generic conservation law of the form:
\begin{equation}
\frac{\partial x}{\partial t} + \nabla \cdot \mathbf{\Gamma} = S
\end{equation}
where $x$ is a conserved quantity, $\mathbf{\Gamma}$ is the flux, and $S$ is a source term, the FVM discretization followed by volume integration and applying the divergence theorem for cell $i$ yields:
\begin{equation}
\label{eq:fvm}
\frac{\partial }{\partial t}(V_i x_i) + (\mathbf{\Gamma}_{i+1/2} \cdot \mathbf{A}_{i+1/2} - \mathbf{\Gamma}_{i-1/2} \cdot \mathbf{A}_{i-1/2})  = V_i S_i, \end{equation}

where:
$x_i$ is the cell-averaged value of $x$ in cell $i$.
$V_i$ is the volume of cell $i$.
$\mathbf{\Gamma}_{i+1/2}$ is the flux at face $i+1/2$.
$\mathbf{A}_{i+1/2}$ is the area of face $i+1/2$. $S_i$ is the cell-averaged source term in cell $i$.

\begin{figure}[hbt]
    \includegraphics[width=1.0\linewidth]{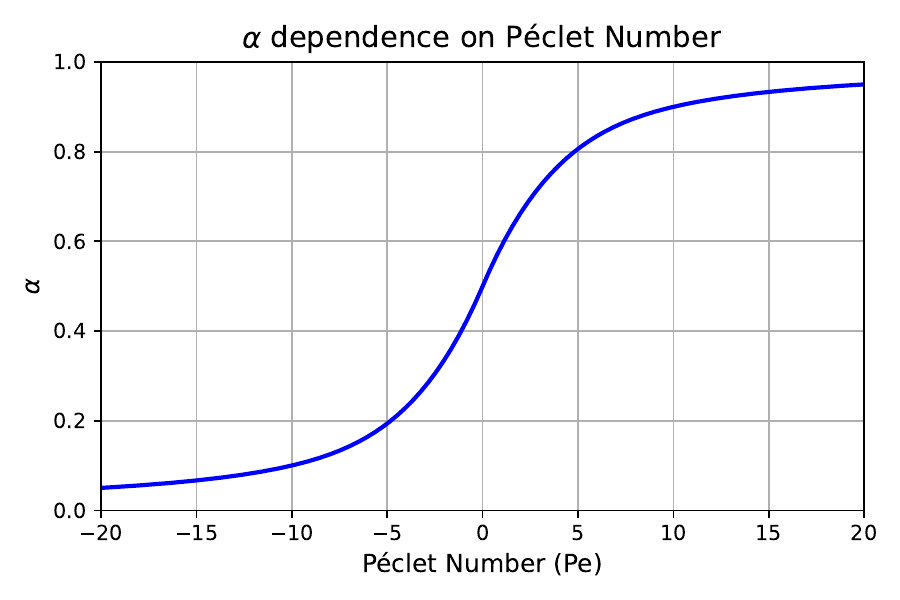}
    \caption{\footnotesize $\alpha$ weighting factor dependence on the P\'{e}clet number in the power-law scheme, as according to equation~\ref{eq:alpha}.}
    \label{fig:peclet}
\end{figure}

Restricting to 1D, the face areas are a constant arbitrary $A$, and the volumes $V_i=d\rnorm_i A$. The area cancels out when dividing equation~\ref{eq:fvm} by $V_i$.
In 1D, the finite volume method reduces to a special case of finite differences:
\begin{equation}
\label{eq:basicPDE}
\frac{\partial }{\partial t}(x_i) + \frac{1}{d \rnorm}({\Gamma}_{i+1/2} - {\Gamma}_{i-1/2})  = S_i, \end{equation}
In general, the fluxes in TORAX are decomposed as
\begin{equation}
\Gamma = -D\frac{\partial x}{\partial \rnorm} + Vx
\end{equation}
where $D$ is a diffusion coefficient and $V$ now denotes a convection coefficient, leading to:
\begin{equation}
\label{eq:fluxes}
\begin{aligned}
\Gamma_{i+1/2} &= -D_{i+1/2}\frac{x_{i+1} - x_{i}}{d\rnorm} + V_{i+1/2}x_{i+1/2} \\
\Gamma_{i-1/2} &= -D_{i-1/2}\frac{x_{i} - x_{i-1}}{d\rnorm} + V_{i-1/2}x_{i-1/2}
\end{aligned}
\end{equation}
The diffusion and convection coefficients are thus calculated on the face grid. 
The value of $x$ on the face grid is approximated by implementing a power-law scheme for P\'{e}clet weighting, which smoothly transitions between central differencing and upwinding, as follows:
\begin{equation}
\label{eq:facevalues}
\begin{aligned}
x_{i+1/2} &= \alpha_{i+1/2}x_i + (1 - \alpha_{i+1/2}) x_{i+1} \\
x_{i-1/2} &= \alpha_{i-1/2}x_i + (1 - \alpha_{i-1/2}) x_{i-1} \\
\end{aligned}
\end{equation}
where the $\alpha$ weighting factor depends on the P\'{e}clet number, defined as:
\begin{equation}
Pe = \frac{V d \rnorm}{D}
\end{equation}
where $V$ is convection and $D$ is diffusion. The power-law scheme is as follows:
\begin{equation}
\label{eq:alpha}
\alpha = \begin{cases}
\frac{Pe - 1}{Pe}  & \text{if } Pe > 10, \\
\frac{(Pe - 1) + (1 - Pe/10)^5}{Pe} & \text{if } 0 < Pe < 10, \\
\frac{(1 + Pe/10)^5 - 1}{Pe} & \text{if } -10 < Pe < 0, \\
-\frac{1}{Pe} & \text{if } Pe < -10.
\end{cases}
\end{equation}
The P\'{e}clet number quantifies the relative strength of convection and diffusion. If the P\'{e}clet number is small and diffusion dominates, then the weighting scheme converges to central differencing. If the absolute value of the P\'{e}clet number is large, and convection dominates, then the scheme converges to upwinding. See figure~\ref{fig:peclet} for an illustration of the $\alpha$ weighting factor dependence on the P\'{e}clet number in the power-law scheme.

Boundary conditions are taken into account by introducing ghost cells $x_{N}$ and $x_{-1}$ whose values are determined by assuming linear extrapolation through the edge cells and the face boundary conditions (for Dirichlet boundary conditions), or by directly satisfying the derivative (Neumann) boundary conditions. These ghost cells are substituted into equation~\ref{eq:facevalues} for the edge values.

Equations~(\ref{eq:fluxes},\ref{eq:facevalues},\ref{eq:alpha}), when combined, define the elements of the discretization matrix and boundary condition vectors for the PDE diffusion term.

\subsection{Solver implementation}

TORAX time integration is based on the theta method, a widely used approach for time discretization of PDEs. 
\subsubsection{Theta method}
The theta method is a weighted average between the explicit and implicit Euler methods. For a generic ODE of the form: 
\begin{equation}
\frac{dx}{dt} = F(x, t)
\end{equation}
where x is the state vector, the theta method approximates the solution at time $t + \Delta t$ as:
\begin{equation}
\label{eq:theta}
x_{t + \Delta t} - x_t = \Delta t \big[ \theta F(x_{t + \Delta t}, t + \Delta t) + (1 - \theta) F(x_t, t)\big]
\end{equation}

where $\theta$ is a user-selected weighting parameter in the range $[0, 1]$.  Different values of $\theta$ correspond to well-known solution methods: explicit Euler ($\theta = 0$), Crank-Nicolson ($\theta = 0.5$), and implicit Euler ($\theta = 1$), which is unconditionally stable. 

\subsubsection{TORAX equation composition}
Upon inspection of the TORAX governing equations ~\ref{eq:Ti}-\ref{eq:psi}, we generalize equation~\ref{eq:theta} and write the TORAX state evolution equation:
\begin{equation}
\label{eq:state_evolution}
\begin{aligned}
& \mathbf{\tilde{T}}(x_{t + \Delta t}, u_{t + \Delta t})\odot\mathbf{x}_{t + \Delta t} - \mathbf{\tilde{T}}(x_t, u_t)\odot\mathbf{x}_t =  \\
& \Delta t \big[ \theta \big( \mathbf{\bar{C}}(x_{t+\Delta t}, u_{t+\Delta t})\mathbf{x}_{t+\Delta t} + \mathbf{c}(x_{t+\Delta t}, u_{t+\Delta t}) \big)  \\ 
& \qquad + (1-\theta) \big( \mathbf{\bar{C}}(x_t, u_t)\mathbf{x}_t + \mathbf{c}(x_{t}, u_{t}) \big) \big]
\end{aligned}
\end{equation}
Starting from an initial condition $\mathbf{x}_0$, equation~\ref{eq:state_evolution} solves for $\mathbf{x}_{t+\Delta t}$ at each timestep. $\mathbf{x}_t$ is the evolving state vector at time $t$, including all variables being solved by the system, and is of length $\#N$, where $\#$ is the number of solved variables. For example, consider a simulation with a gridsize of $25$ solving ion heat transport, electron heat transport, and current diffusion. Then $N=25$, $\#=3$, and $\mathbf{x}_t$ is comprised of $T_i$, $T_e$, and $\psi$, each with its own set of $N$ values, making a total vector length of 75.

$\mathbf{u}_t$ corresponds to all known input parameters at time $t$. This includes boundary conditions, prescribed profiles (e.g. $n_e$ in the example above), and input parameters such as heating powers or locations. 

$\mathbf{\tilde{T}}$ is the transient term (following FiPy nomenclature), where $\odot$ signifies element-wise multiplication. For example, for the $T_e$ equation, $\mathbf{\tilde{T}}=\mathbf{n_e}$, which makes the system nonlinear if $\mathbf{n_e}$ itself is an evolving variable.

$\mathbf{\bar{C}}(x_t, u_t)$ and $\mathbf{\bar{C}}(x_{t+\Delta t}, u_{t+\Delta t})$ are the discretization matrices, of size $\#N\times\#N$. In general, depending on the physics models used, $\mathbf{\bar{C}}$ depends on state variables $\mathbf{x}$, for example through state-variable dependencies of transport coefficients $\chi$, $D$, $V$, plasma conductivity, and ion-electron heat exchange, making the system nonlinear due to the $x_{t+\Delta t}$ dependence. $\mathbf{c}$ is a vector, containing source terms and boundary condition terms.

TORAX provides three solver options for solving equation~\ref{eq:state_evolution}, summarized next.

\subsubsection{Linear solver}
This solver addresses the nonlinearity of equation~\ref{eq:state_evolution} with fixed-point iteration, also known as the predictor-corrector method. For $K$ iterations (user-configurable), an approximation for $\mathbf{x}_{t+\Delta t}$ is obtained by solving the following equation iteratively with $k=1,2,..,K$:
\begin{equation}
\label{eq:predictor_corrector}
\begin{aligned}
& \mathbf{\tilde{T}}(x_{t + \Delta t}^{k-1}, u_{t + \Delta t})\odot\mathbf{x}_{t + \Delta t}^k - \mathbf{\tilde{T}}(x_t, u_t)\odot\mathbf{x}_t =  \\
& \Delta t \big[ \theta \big( \mathbf{\bar{C}}(x_{t+\Delta t}^{k-1}, u_{t+\Delta t})\mathbf{x}_{t+\Delta t}^k + \mathbf{c}(x_{t+\Delta t}^{k-1}, u_{t+\Delta t}) \big)  \\ 
& \qquad + (1-\theta) \big( \mathbf{\bar{C}}(x_t, u_t)\mathbf{x}_t + \mathbf{c}(x_{t}, u_{t}) \big) \big]
\end{aligned}
\end{equation}
and where $\mathbf{x}_{t+\Delta t}^{0} = \mathbf{x}_t$.

By replacing $\mathbf{x}_{t+\Delta t}$ with $\mathbf{x}_{t+\Delta t}^{k-1}$ within the coefficients $\mathbf{\tilde{T}}$, $\mathbf{\bar{C}}$ and $\mathbf{c}$, these coefficients become known at every iteration step, making equation~\ref{eq:predictor_corrector} a \textit{linear} system of equations. $\mathbf{x}_{t+\Delta t}^k$ can then be solved using standard linear algebra methods implemented in JAX.

To further enhance the stability of the linear solver, particularly in the presence of stiff transport coefficients (e.g., when using the QLKNN turbulent transport model, see section~\ref{sec:models}), the Pereverzev-Corrigan method~\cite{pereverzev:2008} is implemented as an option. This method adds a large (user-configurable) artificial diffusion term to the transport equations, balanced by a large inward convection term such that zero extra transport is added at time $t$. These terms stabilize the solution, at the cost of accuracy over short transient phenomena, demanding care in the choice of $\Delta t$ and the value of the artificial diffusion term.

\subsubsection{Newton-Raphson solver}
This solver solves the nonlinear equation~\ref{eq:state_evolution}, using a gradient-based iterative Newton-Raphson root-finding method for finding the value of $\mathbf{x}_{t+\Delta t}$ that renders the residual vector zero:

\begin{equation}
\label{eq:newton_raphson_residual}
\mathbf{R}(\mathbf{x}_{t+\Delta t},\mathbf{x}_t,\mathbf{u}_{t+\Delta t},\mathbf{u}_t, \theta, \Delta t) = 0
\end{equation}
where $\mathbf{R}$ is the LHS-RHS of equation~\ref{eq:state_evolution}.

Starting from an initial guess $\mathbf{x}_{t+\Delta t}=\mathbf{x}_{t+\Delta t}^0$, the Newton-Raphson method linearizes equation~\ref{eq:newton_raphson_residual} about iteration $\mathbf{x}_{t+\Delta t}^k$ and solves the linear system for a step $\delta\mathbf{x}$:
\begin{equation}
\label{eq:newton_raphson_step}
\mathbf{\bar{J}}(\mathbf{x}_{t+\Delta t}^k) \delta\mathbf{x} = -\mathbf{R}(\mathbf{x}_{t+\Delta t}^k)
\end{equation}
where $\mathbf{\bar{J}}$ is the Jacobian of $\mathbf{R}$ with respect to $\mathbf{x}_{t+\Delta t}$. Crucially, JAX automatically calculates $\mathbf{\bar{J}}$ using auto-differentiation.

With $\delta\mathbf{x} = \mathbf{x}_{t+\Delta t}^{k+1} - \mathbf{x}_{t+\Delta t}^{k}$, $\mathbf{x}_{t+\Delta t}^{k+1}$ is solved using standard linear algebra methods implemented in JAX such as LU decomposition. This process iterates until the residual falls below a user-configurable tolerance $\varepsilon$,  i.e: $\| \mathbf{R}(\mathbf{x}_{t+\Delta t}^{k+1}) \|_2 < \varepsilon$, where $\|\cdot\|_2$ is the vector two-norm.

Solver robustness is obtained with a combination of $\delta \mathbf{x}$ line search and $\Delta t$ backtracking. $\delta \mathbf{x}$ line search reduces the step size within a given Newton iteration step, while $\Delta t$ backtracking reduces the overall time step and restarts the entire Newton-Raphson solver for the present timestep, as follows:
\begin{itemize}
    \item If a Newton step leads to an increasing residual, i.e. $\mathbf{R}(\mathbf{x}_{t+\Delta t}^{k+1}) > \mathbf{R}(\mathbf{x}_{t+\Delta t}^k)$, or if $\mathbf{x}_{t+\Delta t}^{k+1}$ is unphysical, e.g. negative temperature, then $\delta \mathbf{x}$ is reduced by a user-configurable factor, and the line-search checks are repeated. The total accumulative reduction factor in a Newton step is denoted $\tau$.
    \item If during the line-search phase, $\tau$ becomes too low, as determined by a user-configurable variable, then the solve is abandoned and $\Delta t$ backtracking is invoked. A new solve attempt is made at a reduced $\Delta t$, reduced by a user-configurable factor, which results in a less nonlinear system.
\end{itemize}

For the initial guess $\mathbf{x}_{t+\Delta t}^0$, two options are available. The user can start from $\mathbf{x}_t$, or use the result of the predictor-corrector linear solver as a warm-start.

\subsubsection{Optimizer solver}
An alternative nonlinear solver using the JAX-compatible jaxopt library~\cite{jaxopt:2021} is also available. This method recasts the residual of equation~\ref{eq:state_evolution} as a loss function, which is minimized using an iterative optimization algorithm. Similar to the Newton-Raphson solver, adaptive timestepping is implemented, where the timestep is reduced if the loss remains above a tolerance at exit. While offering flexibility with different optimization algorithms, this option is relatively untested for TORAX to date.

\subsubsection{Timestep ($\Delta t$) calculation}

TORAX provides two methods for calculating the timestep $\Delta t$, as follows.
\begin{itemize}
\item \textbf{Fixed $\Delta t$}: This method uses a user-configurable constant timestep throughout the simulation. If a nonlinear solver is employed, and adaptive timestepping is enabled, then in practice, some steps may have a lower $\Delta t$ following backtracking.
\item \textbf{Adaptive $\Delta t$}:  This method adapts $\Delta t$ based on the maximum heat conductivity $\chi_{\max}=\max(\chi_i, \chi_e)$.  $\Delta t$ is a multiple of a base timestep inspired by the explicit stability limit for parabolic PDEs:
\begin{equation}
\Delta t_{ \mathrm{base}}=\frac{(d\rnorm)^2}{2\chi_{\max}}
\end{equation}
where $\Delta t = c_{ \mathrm{mult}}^{dt} \Delta t_{ \mathrm{base}}$. $c_{ \mathrm{mult}}^{dt}$ is a user-configurable prefactor.  In practice, $c_{ \mathrm{mult}}^{dt}$ can be significantly larger than unity for implicit solution methods.
\end{itemize}
The adaptive timestep method protects against traversing through fast transients in the simulation, by enforcing $\Delta t \propto 1/\chi$.

\section{Implemented physics models}
\label{sec:models}
TORAX provides a modular framework for incorporating various physics models. This section summarizes the initially implemented models for geometry, transport, sources, and neoclassical physics. Extended physics features are a focus of the short-term development roadmap. Specific plans are summarized following each subsection.

\subsection{Magnetic geometry}
TORAX presently supports two geometry models for determining the metric coefficients and flux surface averaged geometry variables required for the transport equations.
\begin{itemize}
    \item \textbf{CHEASE:} This model utilizes equilibrium data from the CHEASE fixed boundary Grad-Shafranov equilibrium code~\cite{lutjens:1996}. Users provide a CHEASE output file, and TORAX extracts the relevant geometric quantities.
    \item \textbf{Circular:} For testing and demonstration purposes, TORAX includes a simplified circular geometry model. This model assumes a circular plasma cross-section and corrects for elongation to approximate the metric coefficients.
\end{itemize}
Using $\psi$ and the magnetic geometry terms, the toroidal current density is calculated as follows:
\begin{equation}
\label{eq:jtot}
j_\mathrm{tor} = \frac{R_0^2}{8\pi^3\mu_0}\frac{\partial }{\partial V}\left(\frac{g_2 g_3 J}{\rho} \frac{\partial \psi}{\partial \rho}\right)
\end{equation}
where $V$ is the volume enclosed by a flux surface.

The safety factor $q$, a measure of magnetic field line pitch, is calculated as follows:
\begin{equation}
\label{eq:q}
q = 2\pi B_0 \rho \left( \frac{\partial \psi}{\partial \rho} \right)^{-1}
\end{equation}
With $q(\rho=0)=\frac{2B_0}{\mu_0 j_{tot}(\rho=0) R_0}$.



To enable simulations of tokamak scenarios with dynamic equilibra, the TORAX roadmap includes incorporating time-dependent geometry, e.g. by incorporating multiple equilibrium files and interpolating the geometry variables at the times required by the solver. Generalization to geometry data beyond CHEASE is also planned.

\subsection{Plasma composition, initial and prescribed conditions}

Presently, TORAX only accommodates a single main ion species, configured with its atomic mass number ($A_i$) and charge state ($Z_i$). The plasma effective charge, $Z_\textit{eff}$, is assumed to be radially flat and is also user-configurable. A single impurity with charge state $Z_\textit{imp}$ is specified to accommodate $Z_\textit{eff} > 1$. The main ion density dilution is then calculated as follows:
\begin{equation}
n_i=(Z_\textit{imp}-Z_\textit{eff})/(Z_\textit{imp}-1)n_e
\end{equation}
Initial conditions for the evolving profiles $T_i$, $T_e$, $n_e$, and $\psi$ are user-configurable. For $T_{i,e}$, both the initial core ($r=0$) value and the boundary condition at $\rnorm=1$ are provided. Initial $T_{i,e}$ profiles are then a linear interpolation between these values.

For $n_e$, the user specifies a line-averaged density, in either absolute units or scaled to the Greenwald density $n_\mathrm{GW}=\frac{I_p}{\pi a^2}~[10^{20} m^{-3}]$, and the initial peaking factor. The initial $n_e$ profile, including the edge boundary condition, is then a linear function scaled to match the desired line-averaged density and peaking.

If any of the $T_{i,e}$, $n_e$ equations are set to be non-evolving (i.e., not evolved by the PDE stepper), then time-dependent prescribed profiles are user-configurable.

For the poloidal flux $\psi(\hat{\rho})$, the user specifies if the initial condition is based on a prescribed current profile, $j_\mathrm{tor}=j_0(1-\rnorm^2)^\nu$ (with $j_0$ scaled to match $I_p$, and $\nu$ is user-configurable), or from the $\psi$ provided in a CHEASE geometry file. The prescribed current profile option is always used for the circular geometry. The total current $I_p$ can be user-configured or determined by the CHEASE geometry file. In the latter case, the $\psi$ provided by CHEASE can still be used, but is scaled by the ratio of the desired $I_p$ and the CHEASE $I_p$.

In the development roadmap, more flexible initial condition setting is planned, such as from a wider variety of formulas, from arbitrary arrays, or from arbitrary times from existing TORAX output files.

\subsection{Transport models}
Turbulent transport determines the values of the transport coefficients ($\chi_i$, $\chi_e$, $D_e$, $V_e$) in equations~(\ref{eq:Ti}-\ref{eq:ne}), and is a key ingredient in core transport simulations. Theory-based turbulent transport models provide the largest source of nonlinearity in the PDE system. TORAX currently offers three transport models:
\begin{itemize}
\item \textbf{Constant:} This simple model sets all transport coefficients to constant, user-configurable values. While not physically realistic, it can be useful for testing purposes.
\item \textbf{CGM:} The critical gradient model (CGM) is a simple theory-based model, capturing the basic feature of tokamak turbulent transport, critical temperature gradient transport thresholds~\cite{garbet:2004}. The model is a simple way to introduce transport coefficient nonlinearity, and is mainly used for testing purposes. \begin{equation}
\chi_i = \begin{cases}
\chi_{GB} \text{C} (\frac{R}{L_{Ti}} - \frac{R}{L_{Ti,\textit{crit}}})^{\alpha} & \text{if } \frac{R}{L_{Ti}} \ge \frac{R}{L_{Ti,\textit{crit}}} \\
\chi_{min}  & \text{if } \frac{R}{L_{Ti}} < \frac{R}{L_{Ti,\textit{crit}}}
\end{cases}
\end{equation}
with the GyroBohm scaling factor
\begin{equation}
\chi_{GB} = \frac{(A_i m_p)^{1/2}}{(eB_0)^2}\frac{(T_i k_B)^{3/2}}{R_\textit{maj}}
\end{equation}
and the Guo-Romanelli ion-temperature-gradient (ITG) mode critical gradient formula~\cite{guo:1993}
\begin{equation}
R/L_{Ti,crit} = \frac{4}{3}(1 + T_i/T_e)(1 + 2|\hat{s}|/q)
\end{equation}
where $\chi_\textit{min}$ is a user-configurable minimum allowed $\chi$, $L_{Ti}\equiv-\frac{T_i}{\nabla T_i}$ is the ion temperature gradient length, $A_i$ is the main ion atomic mass number, $m_p$ the proton mass, $e$ the electron charge, $B_0$ the magnetic field on axis, and $R_\mathrm{maj}$ the major radius. The stiffness factor $C$ and the exponent $\alpha$ are user-configurable model parameters.

Regarding additional transport coefficient outputs, the electron heat conductivity, $\chi_e$, and particle diffusivity, $D_e$, are scaled to $\chi_i$ using user-configurable model parameters. The particle convection velocity, $V_e$, is user-defined.

\item \textbf{QLKNN:} This is a ML-surrogate model trained on a large dataset of the QuaLiKiz quasilinear gyrokinetic code~\cite{bourdelle:2015,citrin:2017}. Specifically, TORAX presently employs the QLKNN-hyper-10D model (QLKNN10D)~\cite{vandeplassche:2020}, which features a 10D input hypercube and separate NNs for ion-temperature-gradient (ITG), trapped-electron-mode (TEM), and electron-temperature-gradient (ETG) mode turbulent fluxes. The NNs take as input local plasma parameters, such as normalized gradients of temperature and density, temperature ratios, safety factor, magnetic shear, $Z_{eff}$, and normalized collisionality, and outputs turbulent fluxes for ion and electron heat and particle transport. The QLKNN model is significantly faster than direct gyrokinetic simulations, enabling fast and accurate simulation within its range of validity. The ability to seamlessly couple ML-surrogate models is a key TORAX feature. TORAX depends only on the open source weights and biases of the QLKNN model, and includes dedicated JAX inference code written with the Flax library~\cite{flax:2020}.
\end{itemize}

For all transport models, optional spatial smoothing of the transport coefficients using a Gaussian convolution kernel is implemented, to improve solver convergence rates, an issue which can arise with stiff transport coefficients such as from QLKNN. Furthermore, for all transport models, the user can set inner (towards the center) and/or outer (towards the edge) radial zones where the transport coefficients are prescribed to fixed values, overwriting the transport model outputs. This is useful, for example, to model the L-mode edge where QLKNN10D is not validated, or residual transport in the inner core area due to MHD (e.g. sawteeth) or slab or electromagnetic modes which QLKNN10D cannot capture~\cite{kumar:2021}.

\begin{figure}[h]
    \includegraphics[width=1.0\linewidth]{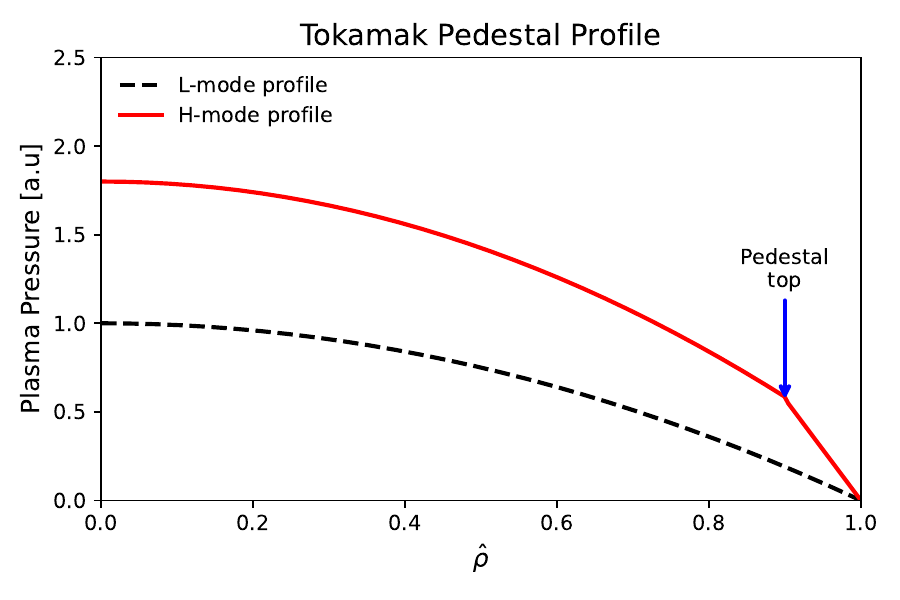}
    \caption{\footnotesize Pedagogical illustration of a tokamak low confinement mode (L-mode) and high confinement mode (H-mode), with the latter characterised by an edge transport barrier with reduced transport, leading to locally high pressure gradients dubbed the "pedestal".}
    \label{fig:pedestal}
\end{figure}

The edge-transport-barrier, or pedestal, is a narrow region near the LCFS where turbulence is suppressed at sufficiently high input power, known as H-mode~\cite{Wagner_PRL82}. See figure~\ref{fig:pedestal} for an illustration. The pedestal effect can be mimicked in TORAX through an adaptive source term which sets a desired value (pedestal height) of $T_e$, $T_i$ and $n_e$, at a user-configurable location (pedestal width).

In the TORAX roadmap, coupling to additional transport models is envisaged, including semi-empirical models such as Bohm/gyroBohm and H-mode confinement scaling law adaptive models, as well as more ML-surrogates of theory-based models, both for core turbulence and pedestal predictions. A more physically consistent approach for setting up the pedestal will be implemented by incorporating adaptive transport coefficients in the pedestal region, as opposed to an adaptive local source/sink term.

\subsection{Neoclassical physics}
TORAX employs the Sauter model~\cite{sauter:1999} to calculate the bootstrap current density, $j_{bs}$, and the neoclassical conductivity, $\sigma_{||}$, used in the current diffusion equation (Eq. \ref{eq:psi}). The Sauter model is a widely-used analytical formulation that provides a relatively fast and differentiable approximation for these neoclassical quantities.

Future work can incorporate more recent neoclassical physics parameterizations~\cite{redl:2021}, and also set neoclassical transport coefficients themselves. This can be of importance for ion heat transport in the inner core. When extending TORAX to include impurity transport, incorporating fast analytical neoclassical models for heavy impurity transport will be of great importance~\cite{angioni:2021,fajardo:2023}.

\subsection{Sources}
The source terms in the governing equations~\ref{eq:Ti}-\ref{eq:psi} are comprised of a summation of individual source/sink terms. Each of these terms can be configured to be either:
\begin{itemize}
\item \textbf{Implicit:} Where needed in the theta-method, the source term is calculated based on the current guess for the state at $t+\Delta t$.
\item \textbf{Explicit:}  The source term is always calculated based on the state of the system at the beginning of the timestep, even if the solver $\theta>0$. This makes the PDE system less nonlinear, avoids the incorporation of the source in the residual Jacobian if solving with Newton-Raphson, and leads to a single source calculation per timestep.
\end{itemize}
Explicit treatment is less accurate, but can be justified and computationally beneficial for sources with complex but slow-evolving physics. Furthermore, explicit source calculations do not need to be JAX-compatible, since explicit sources are an input into the PDE stepper, and do not require JIT compilation (see section~\ref{sec:software}). Conversely, implicit treatment can be important for accurately resolving the impact of fast-evolving source terms. 

All sources can optionally be set to zero, prescribed with non-physics-based formulas (currently Gaussian or exponential) with user-configurable time-dependent parameters like amplitude, width, and location, or calculated with a dedicated physics-based model. Not all sources currently have a model implementation. However, the code modular structure facilitates easy coupling of additional source models in future work. Specifics of source models currently implemented in TORAX follow:

\subsubsection{Ion-electron heat exchange}
The collisional heat exchange power density is calculated as
\begin{equation}
Q_{ei} = \frac{1.5 n_e (T_i - T_e)}{A_i m_p \tau_e},    
\end{equation}
where $A_i$ is the atomic mass number of the main ion species, $m_p$ is the proton mass, and $\tau_e$ is the electron collision time, given by:

\begin{equation}
\tau_e = \frac{12 \pi^{3/2} \epsilon_0^2 m_e^{1/2} (k_B T_e)^{3/2}}{n_e e^4 \ln \Lambda_{ei}}, 
\end{equation}
where $\epsilon_0$ is the permittivity of free space, $m_e$ is the electron mass,
$e$ is the elementary charge, and $\ln \Lambda_{ei}$ is the Coulomb logarithm for electron-ion collisions given by:
\begin{equation}
\ln \Lambda_{ei} = 15.2 - 0.5 \ln \left(\frac{n_e}{10^{20} \text{ m}^{-3}}\right) + \ln (T_e).
\end{equation}
$Q_{ei}$ is added to the electron heat sources, meaning that positive $Q_{ei}$ with $T_i>T_e$ heats the electrons. Conversely, $-Q_{ei}$ is added to the ion heat sources.

\subsubsection{Fusion power}
TORAX optionally calculates the fusion power density assuming a 50-50 deuterium-tritium (D-T) fuel mixture using the Bosch-Hale parameterization~\cite{bosch:1992} for the D-T fusion reactivity $\langle \sigma v \rangle$:
\begin{equation}
P_{fus} = E_{fus} \frac{1}{4} n_i^2 \langle \sigma v \rangle
\end{equation}
where $E_{fus} = 17.6$ MeV is the energy released per fusion reaction, $n_i$ is the ion density, and $\langle \sigma v \rangle$ is given by:
\begin{equation}
\langle \sigma v \rangle = C_1 \theta \sqrt{\frac{\xi}{m_rc^2 T_i^3}} \exp(-3\xi)
\end{equation}
with
\begin{equation}
\theta = \frac{T_i}{1-\frac{T_i (C_2+T_i(C_4+T_iC_6))}{1+T_i(C_3+T_i(C_5+T_i C_7))}}
\end{equation}
and
\begin{equation}
\xi = \left(\frac{B_G^2}{4\theta}\right)^{1/3}    
\end{equation}
where $T_i$ is the ion temperature in keV, $m_rc^2$ is the reduced mass of the D-T pair. The values of $m_rc^2$, the Gamov constant $B_G$, and the constants $C_1$ through $C_7$ are provided in the Bosch-Hale paper.

TORAX partitions the fusion power between ions and electrons using the parameterized alpha particle slowing down model of Mikkelsen~\cite{mikkelsen:1983}, which neglects the slowing down time itself.

\subsubsection{Ohmic power}
The Ohmic power density, $P_\mathrm{ohm}$, arising from resistive dissipation of the plasma current, is calculated as:
\begin{equation}
P_\mathrm{ohm} = \frac{j_\mathrm{tor} }{2 \pi R_\mathrm{maj}}\frac{\partial \psi}{\partial t}
\end{equation}
where $j_\mathrm{tor}$ is the toroidal current density, and $R_\mathrm{maj}$ is the major radius. The loop voltage $\frac{\partial \psi}{\partial t}$ is calculated according to (Eq. \ref{eq:psi}). $P_\mathrm{ohm}$ is then included as a source term in the electron heat transport equation.

\subsubsection{Auxiliary Heating and Current Drive}
While auxiliary heating such as neutral beam injection (NBI), ion cyclotron resonance heating (ICRH), etc, and their associated non-inductive current drives, can all be prescribed with formulas, presently no dedicated physics models are available within TORAX. Future work envisages incorporating more sophisticated physics-based models or ML-surrogate models, enhancing the fidelity of the simulation. For explicit sources, these can also come from external codes (not necessarily JAX compatible) coupled to TORAX in larger workflows. 

Presently, a built-in non-physics-based Gaussian formulation of a generic ion and electron heat source is available in TORAX, with user configurable location, Gaussian width, and fractional heating of ions and electrons.

\subsubsection{Particle Sources}
Similar to auxiliary heating and current drive, particle sources can also be configured using either prescribed formulas. Presently, TORAX provides three built-in formula-based particle sources for the $n_e$ equation:
\begin{itemize}
\item \textbf{Gas Puff:} An exponential function with configurable parameters models the ionization of neutral gas injected from the plasma edge.
\item \textbf{Pellet Injection:} A Gaussian function approximates the deposition of particles from pellets injected into the plasma core. The time-dependent configuration parameter feature allows either a continuous approximation or discrete pellets to be modelled.
\item \textbf{Neutral Beam Injection (NBI):}  A Gaussian function models the ionization of neutral particles injected by a neutral beam.
\end{itemize}
Future work envisages coupling physics-based models and/or ML-surrogates.

\subsubsection{Radiation}
Currently, TORAX does not include dedicated models e.g. for cyclotron radiation, Bremsstrahlung, recombination, or line radiation are yet coupled to TORAX. This is left for future work.

\section{Software design and implementation}
\label{sec:software}
The TORAX codebase is designed with modularity in mind, facilitating extension and maintenance. 
\subsection{Code Structure and Modularity}
The main components of TORAX are organized into distinct modules:
\begin{itemize}
    \setlength\itemsep{0.1em}
    \item Geometry: Handles the representation and manipulation of tokamak geometry, including loading equilibrium data from external files and calculating geometric quantities required for the transport equations.
    \item Transport Models: Implements different transport models for calculating turbulent transport coefficients.
    \item Sources: Defines and manages the various sources and sinks driving the evolution of plasma profiles.
    \item Steppers: Provides numerical solvers for advancing the system of equations in time.
    \item Time Step Calculators: Determines the appropriate time step for each iteration of the solver.
    \item State: Defines data structures for representing the evolving plasma state, and output data structures representing physics and numerics outputs over the simulation time, such as transport coefficients, core profiles, sources, and stepper numerics.
\end{itemize}

The modules include constructor methods and class definitions for Python classes corresponding to simulation physical or numerical components such as \textsf{Geometry}, \textsf{TransportModel}, \textsf{Stepper}, \textsf{SourceProfiles}, which are passed throughout the code. Specific models or solvers are child classes of these objects. For example, a specific turbulent transport model is a child class of \textsf{TransportModel} with a concrete implementation of the call method, ensuring a unified API and shared methods such as the Gaussian smoothing kernel. This aids extensions to new models.

A \textsf{Sim} class, within the sim module, holds all components of the simulation and provides a method that runs the simulation macro time loop and associated glue code. This module is used by the standalone TORAX driver, or can be imported by external workflows to incorporate TORAX into wider simulation frameworks. See Algorithm~\ref{alg:run_simulation} for an overview of the main simulation loop.
\begin{algorithm}[H]
\caption{TORAX simulation procedure (`sim.run\_simulation`)}
\label{alg:run_simulation}
\begin{algorithmic}[1]
\State $t \gets t_{initial}$  \Comment{Initialize simulation time}
\State {Initialize \textsf{geometry}, \textsf{core\_profiles}, \textsf{core\_sources}}
\While{$t < t_{final}$} 
\State {Calculate all \textsf{explicit\_source\_profiles}}
\State $\Delta t \gets \textsf{TimeStepCalculator.next\_dt}(...)$  \Comment{Calculate timestep}
\State {Interpolate all prescribed quantities at $t+\Delta t$}

\State $x_{t+\Delta t} \gets \textsf{SimulationStepFn}(x_t, u_t, u_{t+\Delta t}, \Delta t, ...)$ \Comment{Advance state by one timestep}
\State $t \gets t + \Delta t$   \Comment{Update simulation time}
\State \textsf{torax\_outputs.append(sim\_state})  \Comment{Append state to history for output}
\EndWhile
\end{algorithmic}
\end{algorithm}

In the abstraction above for the \textsf{SimulationStepFn} call, $u$ contains the source runtime parameters, explicit sources (for $u_t$), geometry, prescribed profiles, and boundary conditions, whereas $x$ is the evolving state variables. \textsf{SimulationStepFn} also contains the callables for the PDE stepper, the transport model, and associated runtime parameters. 

\subsection{Simulation I/O}

\subsubsection{Simulation input configuration}
Users specify simulation parameters, such as boundary conditions, initial conditions, heating powers, and solver settings, by providing a Python nested dictionary. The config dictionary provides constructor arguments for constructing the various simulation objects, as well as their runtime parameters, and ultimately the TORAX \textsf{Sim} class. For testing, interactive experimentation, and custom workflows, it is also possible to circumvent the config dictionary and construct the \textsf{Sim} class manually.

TORAX supports time-dependent simulation inputs, enabling the modeling of dynamic tokamak scenarios.  Users can define time series for various parameters on the config level, such as for plasma current, heating power, boundary conditions. TORAX implements either piecewise linear or stepwise interpolation schemes to handle these time-dependent inputs, as per user configuration per input.

\subsubsection{Dynamic vs. Static Parameters}
TORAX differentiates between dynamic and static parameters, affecting the JIT compilation behavior (see section~\ref{sec:JIT}). Dynamic parameters can be modified between simulation runs without triggering recompilation, while static parameters define the fundamental structure of the simulation, and require recompilation if changed. These include the list of evolving variables, the grid size, the stepper functions, and transport models.

\subsubsection{Simulation outputs}
The outputs of the \textsf{run} method of the \textsf{Sim} class are dataclasses analogous to IMAS Interface Data Structures (IDSs), such as core\_profiles, core\_sources, and core\_transport, containing a time history of all associated TORAX variables. In the native TORAX simulation driver, following the completion of the macro time loop, these variables are flattened, converted to an xarray Dataset object~\cite{hoyer:xarray}, and saved to disk as a netCDF file at a user-defined path. A basic visualization module is in place, to browse salient simulation outputs with a time-slider, including comparisons between different simulations simultaneously.

\subsection{JAX JIT compilation and performance}
\label{sec:JIT}
TORAX uses JAX's just-in-time (JIT) compilation capabilities to achieve fast simulation runtimes. JIT compilation transforms Python functions into optimized XLA (Accelerated Linear Algebra) code, enabling efficient execution on CPUs, GPUs, and TPUs. The first call to a JIT-compiled function triggers a tracing and compilation process, which adds to the total simulation computation time. Subsequent calls do not recompile, and execute the optimized XLA code, bypassing the Python interpreter and achieving significant speedups. Compiled functions are cached in memory and can be used in subsequent TORAX simulation calls as long as the process is continued, even with new configs, as long as no static parameters are modified. For new processes, recompilation is necessary. Implementing the JAX persistent on-disk compilation cache in TORAX is currently under development. JIT compilation can be disabled in TORAX by an environment variable, aiding with interactive debugging.

Only the computationally intensive components of a TORAX simulation are compiled. This includes the source models, geometry calculations, neoclassical models, turbulent transport models, matrix manipulations, and the linear algebra solver. For the Newton-Raphson solver, the PDE residual function and its Jacobian are compiled. Crucially, the control flow logic of the iterative solver is not compiled, to avoid compilation inefficiencies, particularly with CPU. This design choice has advantages and disadvantages:

\begin{itemize}
    \item \textbf{Advantages: } 
    \begin{itemize}[label=$\circ$]
        \item Significant compilation speedup (up to factor 10) on CPU.
        \item All control flow and glue code can be programmed in standard Python, which has less constraints. It makes it easier to handle input and output structures, plotting, logging, and coupling external non-JAX models (e.g. as explicit sources, or geometry) to the TORAX solver.
    \end{itemize}
    \item \textbf{Disadvantages: } 
    \begin{itemize}[label=$\circ$]
        \item End-to-end differentiation of a TORAX simulation, e.g. taking the derivative of the final output state with respect to any input parameter, is not possible in a single step. However, the issue is mitigated by the PDE residual Jacobian being compiled and differentiable, making end-to-end differentiation possible via Forward Sensitivity Analysis. See section~\ref{sec:sensitivity}.
        \item Developers need to be aware of the boundaries between elements in the codebase that need to be JAX compatible, and which are not, adding developer overhead.
    \end{itemize}
\end{itemize}

\subsubsection{Computational time for example simulation}
We present a brief summary of simulation compilation and runtime performance, with different solver settings. All runs are based on the \textsf{iterhybrid\_rampup.py} configuration file in the TORAX examples folder. This run evolves 80 seconds of ITER rampup (time-dependent $I_p$), with the QLKNN10D transport model, a gridsize of $N=25$, and simulates ion and electron heat transport, particle transport, and current diffusion. A fixed $\Delta t=2s$ is used. We compare performance when using the Newton-Raphson nonlinear solver, predictor-corrector with 1 corrector step, and predictor-corrector with 10 corrector steps. Simulations were performed on a single compute node equipped with a 48-core AMD EPYC 7B13 processor.

\begin{table}[ht]
\centering
\caption{TORAX Simulation Performance Comparison}
\label{tab:performance}
\footnotesize 
\begin{tabular}{l p{1.8cm}  p{1.8cm}} 
\hline
\textbf{Solver} & \textbf{Compile [s]} & \textbf{Runtime [s]} \\
\hline
Newton-Raphson & 15.6 & 22 \\
Predictor-Corrector (1 step) & 4.5 & 6.5 \\
Predictor-Corrector (10 steps) & 4.6 & 8 \\
\hline
\end{tabular}
\end{table}

Especially considering that compilation overhead is mitigated by in-memory and persistent caching, these runs demonstrate faster-than-realtime simulation capability. This is however not a general conclusion, since resolving transient phenomena on faster timescales, if needed, would slow down the simulation. 

\begin{figure}[hbt]
    \includegraphics[width=1.0\linewidth]{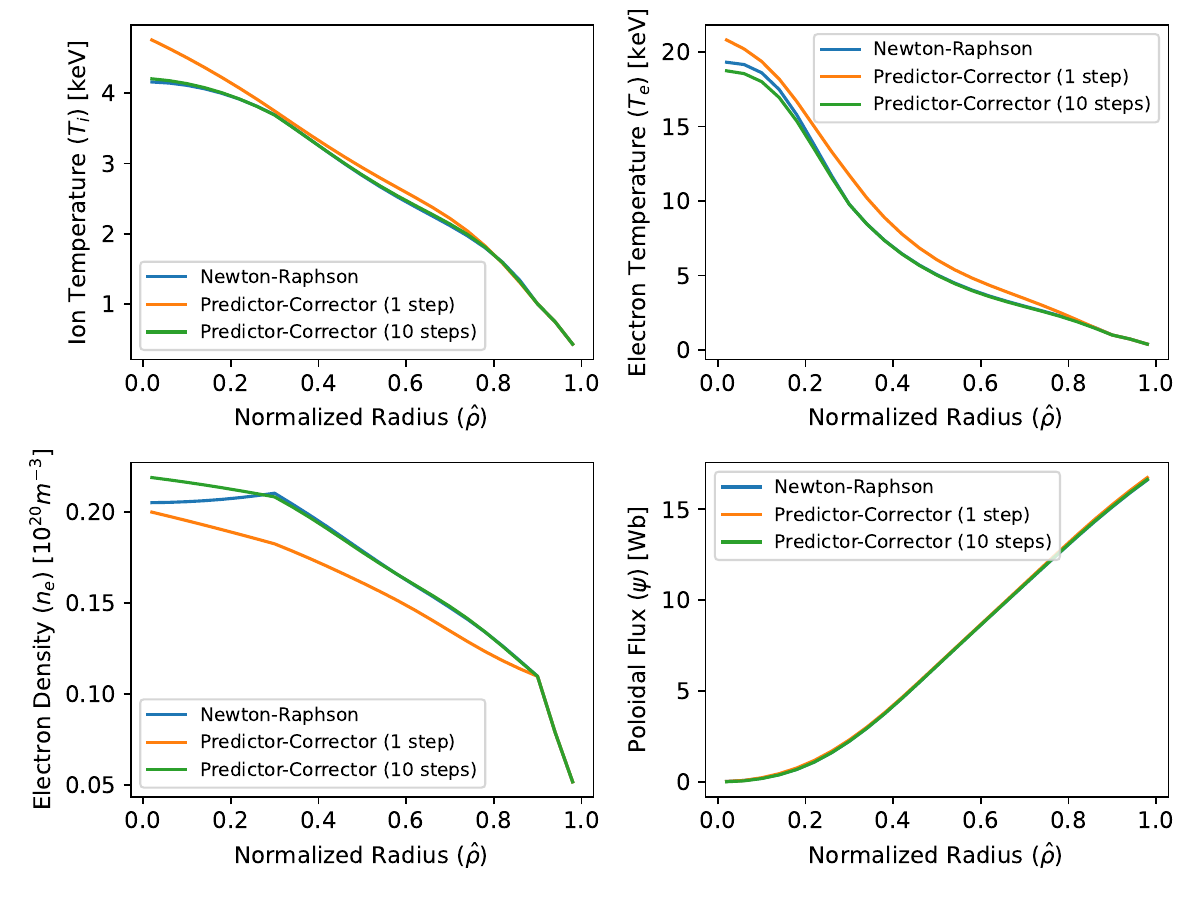}
    \caption{\footnotesize Comparison of simulated $T_i$, $T_e$, $n_e$, and $\psi$ profiles at $t = 10$ s for the Newton-Raphson, Predictor-Corrector (1 step), and Predictor-Corrector (10 steps) solvers, using the example \textsf{iterhybrid\_rampup.py} configuration.}
    \label{fig:solver_comparison}
\end{figure}

Figure~\ref{fig:solver_comparison} compares the three simulations. The 10-step predictor-corrector version converges to the proper nonlinear solution from Newton-Raphson, demonstrating that predictor-corrector can be a robust, fast solver option for use-cases where purely forward simulation is required, without residual and sensitivity information.

\subsubsection{End-to-end differentiation}
\label{sec:sensitivity}
While the TORAX code design does not permit end-to-end direct differentiation, having a differentiable residual enables the application of the Forward Sensitivity Analysis Equation~\cite{cacuci:1981, felici:2011}, to obtain the sensitivity of the state to any input variable. This information can then be used for gradient-based optimization methods~\cite{felici:2012}.

To illustrate this, let us parameterize part of the input vector $u$ with a parameter $p$. For example, a parameterization of the ECRH power amplitude time trace. Following the application of the iterative solver, we obtain a state $x_{t+\Delta t}$ which is a solution of the residual:

\begin{equation}
\label{eq:residual}
\mathbf{R}(\mathbf{x}_{t+\Delta t},\mathbf{x}_t,\mathbf{u}_{t+\Delta t},\mathbf{u}_t) = 0
\end{equation}

We now take the total derivative of the residual with respect to $p$, at timestep $k$.

\begin{equation}
\label{ref:forward}
\begin{aligned}
0 = \frac{d \mathbf{R}_k}{d p} = \frac{\partial \mathbf{R}_k}{\partial \mathbf{x}_{k+1}} \frac{\partial \mathbf{x}_{k+1}}{\partial p} \\ + \frac{\partial \mathbf{R}_k}{\partial \mathbf{x}_k} \frac{\partial \mathbf{x}_k}{\partial p} +  \frac{\partial \mathbf{R}_k}{\partial \mathbf{u}_k} \frac{\partial \mathbf{u}_k}{\partial p} & + \frac{\partial \mathbf{R}_k}{\partial \mathbf{u}_{k+1}} \frac{\partial \mathbf{u}_{k+1}}{\partial p}
\end{aligned}
\end{equation}

We are interested in calculating $\frac{\partial \mathbf{x}_{k+1}}{\partial p}$, the sensitivity of the state at time $t+\Delta t$ to arbitrary input parameter. The key point, is that all the $\mathbf{R}$ derivatives can be calculated by JAX in a generalized Jacobian, using auto-differentiation. All $\mathbf{u}$ derivatives can also calculated with JAX gradients. 

Thus, starting from the initial condition $\frac{\partial \mathbf{x}_{0}}{\partial p}$, which is known, equation~\ref{ref:forward} is recursively solved until time $k+1$. This method is not yet implemented in TORAX, but is on the short-term roadmap.

\begin{figure*}[hbt]
    \centering
    \includegraphics[width=1.0\textwidth]{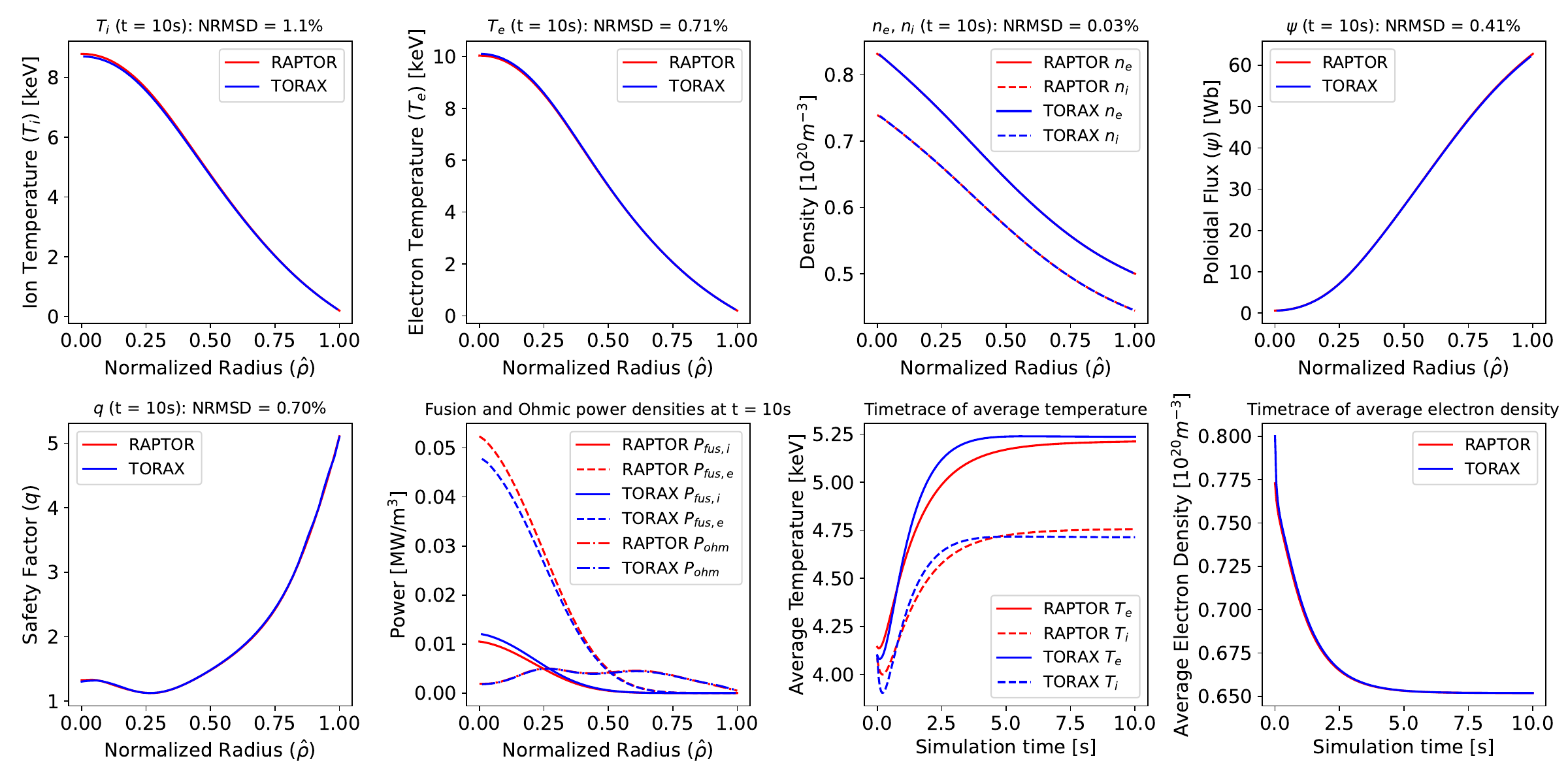}
    \caption{\footnotesize Comparison of simulated profiles from TORAX and RAPTOR for an ITER L-mode scenario. The simulation includes ion and electron heat transport, particle transport, and current diffusion, using the constant transport coefficient model. The top row compares the ion and electron temperatures, electron density, and poloidal flux at the end of the simulation. The normalized root-mean-square deviation (NRMSD) between the TORAX and RAPTOR profiles, interpolated onto the same grid, is indicated in the title of each subplot, demonstrating excellent agreement between the two codes. The bottom row shows the safety factor ($q$-profile), and outputs from the Ohmic and fusion power models, at the end of the simulation. Minor differences in fusion power are attributed to the different models used for fractional power deposition to ions and electrons. The final panels in the bottom row show time traces of line-averaged electron and ion temperatures, and electron density. The temperature profiles have a maximum transient deviation of $\sim2.5\%$.}
    \label{fig:verification1}
\end{figure*}

\section{Verification against RAPTOR}

To validate TORAX, we conducted benchmark simulations against the established RAPTOR code~\cite{felici:2012, felici:2018}, which has itself been validated against other transport solvers \cite{na:2019, vandeplassche:2020}. 

The first comparison is for an ITER-like L-mode scenario, with constant plasma current of $I_p=11.5~MA$, constant external power of $P_{tot}=50~MW$ equally split between ions and electrons, and using the default TORAX CHEASE ITER hybrid scenario equilibrium, as well as its $\psi$ profile for the initial condition, scaled to match our requested $I_p$. Ion and electron heat transport, particle transport, and current diffusion were all modelled, for 10 seconds until stationary state for heat and particle transport, with $\Delta t=0.05~s$. The constant transport model was used with $\chi_i/\chi_e=2$, with inward convection included for the particle transport, and a broad particle source. Fusion power, bootstrap current, Ohmic power, and ion-electron heat exchange were all included in the simulation. The Newton-Raphson solver was used in TORAX. The TORAX configuration file is shown in table~\ref{lst:torax_verification_config}.

Figure~\ref{fig:verification1} shows a comparison between the simulated profiles from TORAX and RAPTOR. For quantitative assessment of the agreement, we calculated the normalized root-mean-square deviation (NRMSD) for each profile at $t = 10$ s, interpolated onto the same grid, using the formula:

\begin{equation}
\text{NRMSD} = \frac{\sqrt{\sum_{i=1}^{N} (y_{i,TORAX} - y_{i,RAPTOR})^2 / N}}{\sum_{i=1}^{N} (y_{i,TORAX})/N} \times 100\%
\end{equation}

where $y$ represents the profile being compared (e.g., $T_i$, $T_e$, $n_e$), and $N$ is the number of radial grid points.

The comparisons show excellent $\sim1\%$ agreement at stationary state. Minor differences in fusion power at stationary state are due to the different models used for the fractional power deposition to ions and electrons. There is a maximum deviation of $\sim2.5\%$ between the temperature profiles during the transient phase.

\begin{figure}[hbt]
    \centering
    \includegraphics[width=0.5\textwidth]{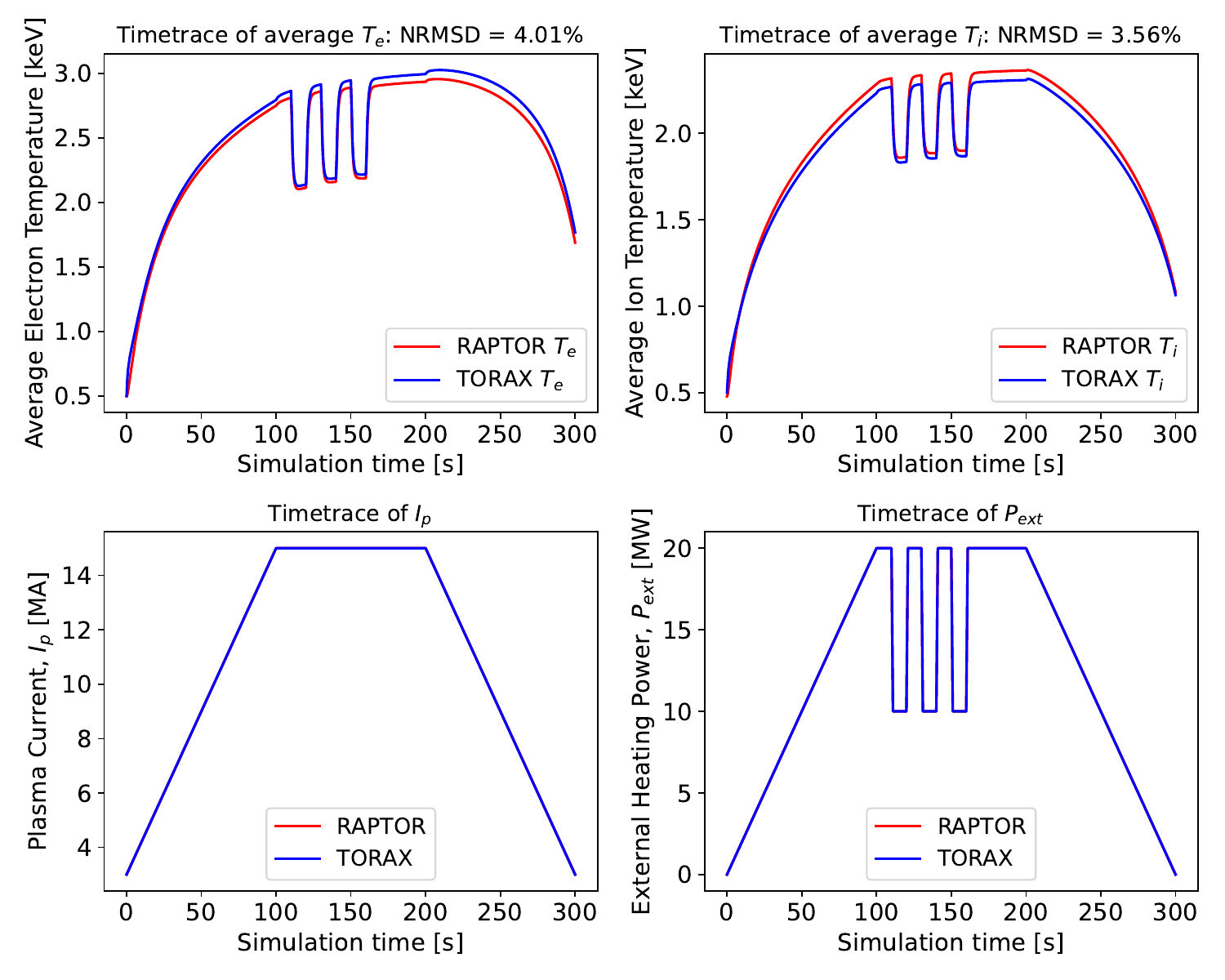}
    \caption{\footnotesize Comparison of simulated profiles from TORAX and RAPTOR for an ITER L-mode scenario with varying current and input power. Constant transport coefficients are used, and the simulation includes ion and electron heat transport, and current diffusion, with a prescribed varying flat density profile set at a Greenwald fraction of 0.5. The bottom row illustrates the input plasma current and external electron heating power, identical for the two codes. Modulation in the heating power at current flattop induces additional dynamics. The top row shows timetrace comparisons for ion and electron temperature. $NRMSD\approx4\%$, with excellent agreement on the timescale dynamics over both energy confinement and current diffusion timescales.}
    \label{fig:verification3}
\end{figure}

A further comparison was made to verify time-dependent inputs. A dynamic ITER L-mode scenario was simulated, using constant transport coefficients. Heat transport and current diffusion were modelled over 300 seconds of plasma evolution with 3 phases: a 100 second current rampup from $3~MA$ to $15~MA$, a 100 second flattop phase, and a 100 second current rampdown back to $3~MA$. 20~MW of pure electron heating was applied, ramping up and down during the current ramps. During the flattop phase, the power was modulated for 60 seconds, to trigger further dynamics. Plasma density was prescribed as flat, with a varying magnitude fixed to a Greenwald fraction of 0.5. The simulation is shown in figure~\ref{fig:verification3}. RAPTOR and TORAX track the dynamics closely, with a NRMSD$<5\%$ for both $T_e$ and $T_i$. The energy confinement timescale is on the order of seconds, evident by the fast (on discharge timescale) relaxation of the temperature profiles during the heating modulation. A longer timescale temperature increase is evident during flattop, which is the current diffusion timescale, leading to increased peaking of the Ohmic heating profile over time, and impacting the temperatures due to the constant transport coefficients.

\section{Conclusions and future work}
TORAX is a new open-source differentiable tokamak core transport simulator, offering significant advantages for fast and accurate tokamak scenario modeling, pulse design, and control. The choice of Python and the JAX library provides a flexible framework for implementing ML-surrogates of physics models, coupling within larger workflows, and fast simulation execution through compilation. TORAX is verified through comparison to the RAPTOR code, demonstrating good agreement in simulated plasma profiles, and any differences generally understood to arise from their different discretization and model implementations.

A key feature in TORAX is a differentiable PDE residual, enabling gradient-based nonlinear PDE solvers, and the use of the Forward Sensitivity Analysis Equation (see Eq.~\ref{ref:forward}) to obtain the sensitivity of the state to arbitrary input parameters, facilitating gradient-based optimization workflows.

Upcoming technical development plans include:
\begin{itemize}
    \item Time dependent geometry
    \item Increased flexibility for setting initial and prescribed conditions
    \item Implementation of forward sensitivity calculations w.r.t. control inputs and parameters
    \item Implementation of a persistent compilation cache
    \item Stationary-state solver
    \item Coupling to IMAS
\end{itemize}

The roadmap foresees the implementation of the following physics models:
\begin{itemize}
    \item MHD models (e.g. sawteeth, neoclassical tearing modes)
    \item Radiation sinks (cyclotron, Bremsstrahlung, line radiation)
    \item Widened range of semi-empirical and theory-based ML-surrogates of turbulent transport models.
    \item Neoclassical transport and multi-ion transport, with a focus on heavy impurities
    \item Momentum transport
\end{itemize}

Contributions in line with the roadmap are welcome. In particular, TORAX is envisaged as a natural framework for coupling of various ML-surrogates of physics models. These could include surrogates for turbulent transport, neoclassical transport, heat and particle sources, line radiation, pedestal physics, core-edge integration, and MHD, among others.

Through its flexibility, speed, and ease of coupling ML-surrogates, the continued development of TORAX targets beyond-state-of-the-art applications for tokamak optimization, pulse-planning, and controller-design workflows.

\section{Acknowledgements}
The TORAX authors thank Olivier Sauter and Antoine Merle for aiding with the provision of CHEASE equilibria, and Allen Wang for early contributions to the open source TORAX codebase.

\newpage  

\begin{table*}[t]  
\centering
\caption{TORAX configuration (Python dict) for the verification benchmark shown in figure~\ref{fig:verification1}.}
\label{lst:torax_verification_config}
\footnotesize
\begin{lstlisting}[language=Python]
CONFIG = {
    'runtime_params': {
        'plasma_composition': {
            'Ai': 2.5,
            'Zeff': 2.0,
            'Zimp': 10,
        },
        'profile_conditions': {
            'Ip': 11.5,
            'Ti_bound_left': 8,
            'Ti_bound_right': 0.2,
            'Te_bound_left': 8,
            'Te_bound_right': 0.2,
            'ne_bound_right': 0.5,
            'nbar_is_fGW': False,
            'nbar': 0.8,
            'npeak': 1.0,
            'set_pedestal': False,
        },
        'numerics': {
            't_final': 10.0,
            'exact_t_final': True,            
            'fixed_dt': 0.05,
            'ion_heat_eq': True,
            'el_heat_eq': True,
            'current_eq': True,
            'dens_eq': True,
        },
    },
    'geometry': {
        'nr': 50,
        'geometry_type': 'chease',
        'geometry_file': 'ITER_hybrid_citrin_equil_cheasedata.mat2cols',
        'Ip_from_parameters': True,
        'Rmaj': 6.2,
        'Rmin': 2.0,
        'B0': 5.3,
    },
    'sources': {
        'j_bootstrap': {},
        'nbi_particle_source': {
            'S_nbi_tot': 3e+21,
            'nbi_deposition_location': 0.2,
            'nbi_particle_width': 0.25,
        },
        'generic_ion_el_heat_source': {
            'rsource': 0.11,
            'w': 0.2,
            'Ptot': 50e6,
            'el_heat_fraction': 0.5,
        },
        'ohmic_heat_source': {},
        'fusion_heat_source': {},
        'qei_source': {},
    },
    'transport': {
        'transport_model': 'constant',
        'constant_params': {
            'chii_const': 2.0,
            'chie_const': 1.0,
            'De_const': 1.0,
            'Ve_const': -0.15,
        },
    },
    'stepper': {
        'stepper_type': 'newton_raphson',
        'predictor_corrector': True,
        'corrector_steps': 5,
    },
    'time_step_calculator': {
        'calculator_type': 'fixed',
    },
}
\end{lstlisting}
\end{table*}

\appendix 

\section{Time dependent geometry impact on convective and source terms}

\label{appendix:phibdot}

For ion and electron heat transport, suppressing the species subscript and multiplying the equation by $V'$, we obtain

\begin{multline}
\label{eq:heattransport}
\frac{3}{2} V'^{-2/3} \left(\frac{\partial }{\partial t}-\frac{\phibdot}{2\phib}\frac{\partial}{\partial\rnorm}\rnorm\right)\left[V'^{5/3} n T\right] = \\ \frac{\partial}{\partial \rnorm} \left[\chi n \frac{g_1}{V'} \frac{\partial T}{\partial \rnorm} - g_0q^{\mathrm{conv}}T\right] + V'Q
\end{multline}

Moving the $\phibdot$ term to the RHS, we obtain:

\begin{multline}
\label{eq:heattransport}
\frac{3}{2} V'^{-2/3} \frac{\partial }{\partial t}\left(V'^{5/3} n T\right) = \frac{3}{4}V'^{-2/3}\frac{\phibdot }{\phib}\frac{\partial}{\partial\rnorm}\rnorm V'^{5/3} n T \\ + \frac{\partial}{\partial \rnorm } \left[\chi n \frac{g_1}{V'} \frac{\partial T}{\partial \rnorm} - g_0q^{\mathrm{conv}}T\right] + V'Q
\end{multline}

Utilizing the identity $a \frac{\partial (b T) }{\partial \rnorm} = \frac{\partial (a b T) }{\partial \rnorm} - b T\frac{\partial a }{\partial \rnorm}$, and rearranging terms, we obtain:

\begin{multline}
\label{eq:heattransport}
\frac{3}{2} V'^{-2/3} \frac{\partial }{\partial t}\left(V'^{5/3} n T\right) = \\ \frac{\partial}{\partial \rnorm } \left[\chi n \frac{g_1}{V'} \frac{\partial T}{\partial \rnorm} - \left(g_0q^{\mathrm{conv}} - \frac{3}{4}\frac{\phibdot }{\phib}\rnorm V' n \right)T \right] + \\ V'Q + \frac{1}{2}V''\frac{\phibdot }{\phib}\rnorm n T 
\end{multline}

Non-zero $\phibdot$ thus results in an effective convection term equal to -$\frac{3}{4}\frac{\phibdot }{\phib}\rnorm V' n$, and an effective additional source term equal to $\frac{1}{2}V''\frac{\phibdot }{\phib}\rnorm n T$, where $V''$ is the second derivative of the plasma volume with respect to $\rnorm$.

For particle transport, using similar considerations, we obtain the following governing equation:

\begin{multline}
\label{eq:ne_rearranged}
\frac{\partial}{\partial t}\left( n_e V' \right) = \\ \frac{\partial}{\partial \rnorm} \left[D_e n_e \frac{g_1}{V'} \frac{\partial n_e}{\partial \rnorm} - \left( g_0V_e - \frac{\phibdot}{2\phib}\rnorm V'\right) n_e\right] + V'S_n
\end{multline}   

where the $\phibdot$ terms leads to an additional effective convection term of -$\frac{\phibdot}{2\phib}\rnorm V'$, and no additional effective source terms.

For current diffusion, we obtain:

\begin{multline}
    \label{eq:psi_rearranged}
 \frac{16 \pi^2 \sigma_{||}\mu_0 \rnorm \phib^2}{F^2}\frac{\partial \psi}{\partial t}-\frac{\rnorm\phibdot}{2\phib}\frac{\partial \psi}{\partial \rnorm}  = \\ \frac{\partial}{\partial \rnorm} \left( \frac{g_2 g_3}{\rnorm} \frac{\partial \psi}{\partial \rnorm} +8 \pi^2 \mu_0 \phib\phibdot\frac{\sigma_{||}\rnorm^2}{F^2}\psi\right) - \\ \frac{8\pi^2 V' \mu_0 \phib}{F^2} \langle \mathbf{B} \cdot \mathbf{j}_{ni} \rangle - 8 \pi^2 \mu_0 \phib\phibdot\psi\frac{\partial}{\partial \rnorm}\left(\frac{\sigma_{||}\rnorm^2}{F^2}\right)
\end{multline} 

where the $\phibdot$ terms leads to an additional effective convection term of - $8 \pi^2 \mu_0 \phib\phibdot\frac{\sigma_{||}\rnorm^2}{F^2}$, and an additional effective source term of $- 8 \pi^2 \mu_0 \phib\phibdot\psi\frac{\partial}{\partial \rnorm}\left(\frac{\sigma_{||}\rnorm^2}{F^2}\right)$.

\bibliography{refs}

\end{document}